\documentclass[preprint2]{aastex}
\usepackage{emulateapj5}
\usepackage{onecolfloat5}

\shortauthors{Chatterjee \& Cordes}
\shorttitle{An Evolving Neutron Star Bow~Shock}
\slugcomment{For the Astrophysical Journal Letters}

\newcommand{\Halpha}{H$\alpha$}
\newcommand{\Edot}{$\dot{E}$}

\begin{document}
\twocolumn[
\title{Smashing the Guitar: An Evolving Neutron Star Bow~Shock}
\author{S. Chatterjee \& J. M. Cordes}
\affil{Department of Astronomy, Cornell University, Ithaca, NY 14853}
\email{shami@astro.cornell.edu, cordes@astro.cornell.edu} 

\begin{abstract}
  The Guitar nebula is a spectacular example of an \Halpha\ bow shock
  nebula produced by the interaction of a neutron star with its
  environment.  The radio pulsar B2224+65 is traveling at
  $\sim$800--1600~km~s$^{-1}$ (for a distance of 1--2~kpc), placing it
  on the high-velocity tail of the pulsar velocity distribution.  Here
  we report time evolution in the shape of the Guitar nebula, the
  first such observations for a bow shock nebula, as seen in \Halpha\
  imaging with the Hubble Space Telescope.  The morphology of the
  nebula provides no evidence for anisotropy in the pulsar wind, nor
  for fluctuations in the pulsar wind luminosity.  The nebula shows
  morphological changes over two epochs spaced by seven years that
  imply the existence of significant gradients and inhomogeneities in
  the ambient interstellar medium.  These observations offer
  astrophysically unique, {\em in situ} probes of length scales
  between $5\times10^{-4}$~pc and $0.012$~pc.  Model fitting suggests
  that the nebula axis --- and thus the three-dimensional velocity
  vector --- lies within 20\arcdeg of the plane of the sky, and also
  jointly constrains the distance to the neutron star and the ambient
  density.
\end{abstract}

]

\section{Introduction}

While the steady decay of spin rates observed in radio pulsars
provides a good estimate of the rate of energy loss \Edot, only a
small fraction of the energy output of a neutron star is typically
converted to directly detectable electromagnetic radiation. Most of
the spindown energy loss is carried away by a relativistic wind, the
properties of which are largely unknown.  The best constraints on the
relativistic wind derive from its interaction with the interstellar
medium (ISM).  This interaction has been observed, for example, in
synchrotron nebulae such as the Crab Nebula \citep{KC84,GA94,MM96},
where wisp structures are observed moving at $\sim 0.5c$
\citep{HMB+02}, as well as bow shocks observed in \Halpha\ emission
from shock-excited neutral gas around PSR~B1957+20 \citep{KH88} and
various other pulsars \citep{CC02,GJS02}.

The Guitar nebula was discovered in deep \Halpha\ imaging observations
with the 5-m Hale telescope at Palomar Observatory \citep{CRL93}.
Using a model for the Galactic electron density distribution
\citep{CL02}, the dispersion measure (DM $= \int_0^Dn_e\,ds = $ 35.3
pc~cm$^{-3}$) of PSR~B2224+65 implies a distance of 1.9~kpc (hereafter
parameterized as D$_{1.9}$).  While the pulsar has a modest \Edot =
10$^{33.1}$ erg~s$^{-1}$, the large space velocity of the pulsar
($\sim 1640$ D$_{1.9}$ km~s$^{-1}$) provides the ram pressure needed
to create a detectable bow shock nebula.  The existence of an
observable bow shock nebula also implies the presence of a significant
neutral hydrogen component in the ISM near the pulsar.  Here we
describe Hubble Space Telescope (HST) observations of the time
evolution of the nebula, and discuss the implications of the
observations for our understanding of neutron star (NS) relativistic
winds and the ISM.

\section{HST Observations and Modeling}

\begin{figure*}[ht]
\epsscale{1.5}
\plotone{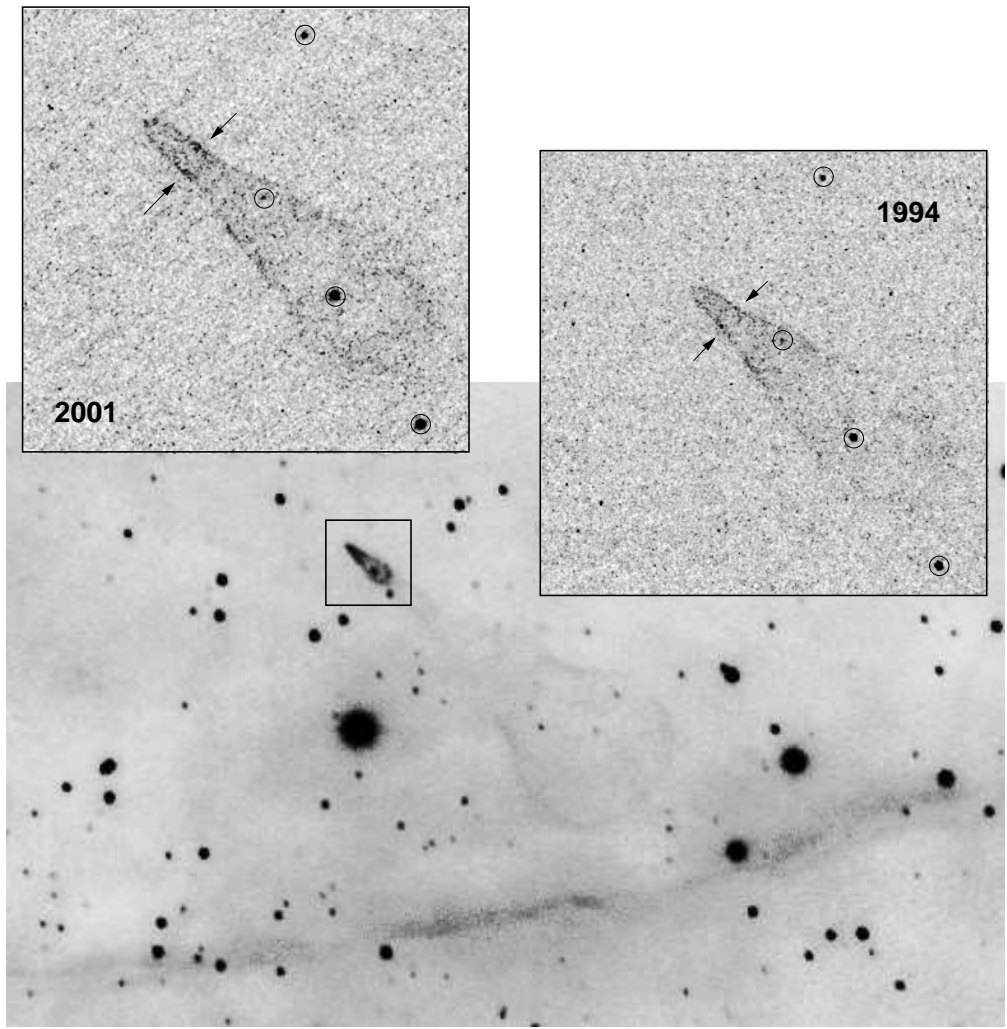}
\caption{\Halpha\ images of the head of the Guitar nebula. The bottom
  panel shows a wide-field image of the Guitar nebula obtained at the
  5-m Hale Telescope at Palomar \citep{CC02}. High resolution images of
  the region marked with a box ($\sim 16$\arcsec\ in size) were
  obtained with the HST PC in 1994 (right) and 2001 (left). North is
  upward and east is to the left.
  Stars are circled, and arrows mark constricted regions in the
  limb-brightened \Halpha\ emission that are conspicuously brighter
  (see text). Note that the neutron star is located at the very tip 
  of the nebula at each epoch.
\label{fig-hst}
}
\end{figure*}

\begin{figure}[t]
\epsscale{1.0}
\plotone{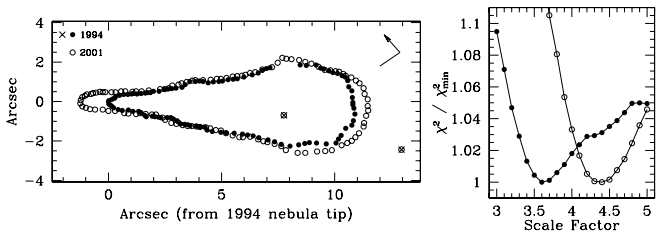}
\caption{
(a) Left: The outline of the limb-brightened head of the Guitar nebula
  in 1994 (filled circles) and 2001 (open circles), obtained by
  eye. The images are aligned using 8 bright stars in the PC image at
  each epoch: to demonstrate the image registration, two stars in the
  PC image section displayed in Figure \ref{fig-hst} are marked with
  crosses in 1994 and open circles in 2001.
(b) Right: Slices through the best fit point on the $\chi^2$ surface
  for the model fit procedure on 1994 data (filled circles) and 2001
  data (open circles), showing the relative change in $\chi^2$ as a
  function of the scale factor $S = D^{-2}_{\rm kpc}
  n_{A}^{-1/2}$.  The implied change in ambient density at the nebula
  tip is $n_A ({\rm 2001}) / n_A ({\rm 1994}) \approx 0.7$.
\label{fig-edgfit}
}
\end{figure}

High resolution HST observations obtained in 1994 December with the
Wide Field and Planetary Camera 2 \citep[WFPC2;][]{wfpc2} have
been described previously \citep{CC02}. 
New WFPC2 observations were obtained in 2001 December,
with about 2.4 times the exposure time of the earlier epoch.
At both epochs, exposures were combined using
variable-pixel linear reconstruction \citep[``Drizzling'';][]{drizzle},
yielding an effective pixel scale 0\farcs0226, roughly half the size
of the 0\farcs0455 pixels on the WFPC2 Planetary Camera chip.
The \Halpha\ images at the two epochs (Figure~\ref{fig-hst}) were
aligned to sub-pixel accuracy using eight stars
(Figure~\ref{fig-edgfit}a).  The tip of the nebula has moved
1.3\arcsec\ in 7 years, consistent with the radio proper motion of
PSR~B2224+65 \citep[$\mu = 182 \pm 3$ mas yr$^{-1}$, at a position
angle $52.1\arcdeg \pm 0.9\arcdeg$;][]{HLA93}. However, unlike the
simple translation with constant shape expected for a bow shock in a
uniform ambient medium, the head of the nebula has changed
morphologically, showing corrugations in the limb-brightened edge as
well as gaps in the \Halpha\ flux, both of which vary with time.

At the tip of the nebula, the stand-off radius of the bow shock 
has increased over the seven-year period, as determined by model
fitting (discussed below) at each epoch.
Pressure balance between the relativistic NS wind and ram pressure
from the ambient medium occurs at the stand-off radius,
\begin{equation}
R_0 = (\dot{E} / 4 \pi c n_A m v_{\ast}^2 )^{1/2},
\label{Eqn:R0}
\end{equation}
where
$n_A$ is the number density of particles with mean mass $m$ and
$v_{\ast}$ is the pulsar velocity.  For the Guitar nebula,
incorporating measured values for the pulsar \Edot\ and proper motion,
the stand-off radius is $R_0 \approx 18 \; {\rm D}_{1.9}^{-1}
n_A^{-1/2}$ astronomical units (AU), and the predicted stand-off angle
is $ \theta_0 = 9.4 \; {\rm D}^{-2}_{1.9} n_A^{-1/2}$ milliarcseconds.
Bright patches appear along the edges of the nebula where it is
constricted, as marked with arrows in Figure \ref{fig-hst}: the
location of this brightening has
moved $\sim 0.8$\arcsec\ parallel to the pulsar motion along both
sides of the nebula. However, in Figure \ref{fig-edgfit}(a), the edges
perpendicular to the nebula axis have moved outwards by less than
0.3\arcsec, remaining essentially static at several points. Meanwhile,
the rear edge of H$\alpha$\ emission has actually moved backwards by
$\sim 0.7$\arcsec\ in 2001 compared to 1994.  Possible explanations
include turbulence in the shocked layers, their interaction with a
complex magnetic field, or changes in any of the variables in the
expression for $R_0$ (Equation~\ref{Eqn:R0}), including \Edot.
However, even for the largest observed glitches \citep{HLJ+02}, 
\Edot\ changes temporarily by $\lesssim 3\%$, a negligible variation
compared to the large morphological changes observed.  Besides being
contrary to our general understanding of spin down, a time-variable
\Edot\ also fails to explain the brightening of the nebula at the
locations where it appears constricted. Indeed, it is difficult to
explain the evolving morphology without invoking variations in the
ambient interstellar density, although instabilities in the shock
structure are also possible \citep[e.g.][]{DM93}. We propose that the
observations reflect the motion of PSR~B2224+65 through random density
inhomogeneities combined with a gradient towards a region of lower
density.  The bright rear edge ($\sim 11\arcsec$ downstream of the
nebula tip in 1994) marks a sharp increase in density which the
NS broke through $\sim 70$ years ago. Presently, the shock
at the rear confines the relativistic NS wind, leading to a
brightening of the head of the nebula and preventing the wind from
powering the rest of the Guitar body, thus producing the dim,
elongated and narrow ``neck'' of the Guitar in Figure~\ref{fig-hst}.

We argue further that the morphology of the head of the Guitar nebula
is an analog for confinement by another high-density
region $\sim 300$ years ago, which created the rounded end of the Guitar
body. Fluctuations in the ambient density cause the body to be
brighter where it appears constricted. In a few hundred years, as the
larger Guitar body fades from view, what is currently the head of the
nebula may expand to become another guitar-like structure. 
The scenario described here, while plausible, needs to
be verified with future high-resolution monitoring observations as
well as time-dependent hydrodynamic modeling of shock fronts
in an ambient medium with significant density fluctuations.

\begin{deluxetable}{lcccccl}
\tablecolumns{5}
\tablewidth{0pc} 
\tablecaption{Best-fit Model Parameters for HST Images\label{table-model}}
\tablehead{ 
\colhead{} & \colhead{Position Angle} & \colhead{Inclination\tablenotemark{\dag}} & 
\colhead{Scale Factor} & \colhead{$\theta_{0}$} \\
\colhead{Epoch} & \colhead{($\arcdeg$)} & \colhead{($\arcdeg$)} &
\colhead{($D^{-2}_{\rm kpc} n_{A}^{-1/2}$)} & \colhead{(\arcsec)} 
}
\startdata
1994 & $48 \pm 2$ & $90 \pm 30$ & $3.6 \pm 1.2$ & $0.12 \pm 0.04$\\
2001 & $50 \pm 2$ & $90 \pm 20$ & $4.4 \pm 1.2$ & $0.15 \pm 0.04$ 
\enddata
\tablenotetext{\dag}{The fit for inclination angle was inconclusive,
especially for the 1994 data, as discussed in \citet{CC02}.}
\end{deluxetable}

Currently, in order to quantify the change in stand-off radius between
epochs, we have modeled the shock front under the assumption that the
nebula tip is in quasi-static equilibrium with the ambient medium.  A
momentum-conserving bow shock model \citep{W96} was adapted to fit the
\Halpha\ emission at each epoch. The momentum-conserving description
has known limitations \citep{BB01}, especially since it applies only
to an ambient medium of uniform density. To avoid these problems, we
restricted the model fit to within 2\farcs6 of the tip of the nebula,
where it is smooth and symmetric.  The model is parameterized by the
position angle and the inclination of the nebula to the line of sight
(neither of which are expected to change significantly between
epochs), the thickness of the shocked layer that emits \Halpha, and by
a scale factor $S = D^{-2}_{\rm kpc} n_{A}^{-1/2}$,
which isolates the dependence of the apparent size of the nebula on
distance and ambient density. Details of the model fitting procedure
(for the 1994 data) are given in \citet{CC02} and the best-fit
parameters for both epochs (position angle, inclination, scale factor
and stand-off angle) are listed in Table~\ref{table-model}.  At both
epochs, the fit constrains the nebula to lie in the plane of the sky.
As shown in Figure \ref{fig-edgfit}(b), the best-fit scale factor
varies significantly between 1994 and 2001. Since the fractional
change in distance to the nebula is negligible over seven years, the
change in scale factor implies a decrease in ambient density by a
factor $\sim 0.7$ (from 0.006~cm$^{-3}$ to 0.004~cm$^{-3}$ for D =
1.9~kpc) over 1.3\arcsec, corresponding to a length scale of
$2500\,{\rm D}_{1.9}$~AU.
The implied change in DM for density changes on this length scale is
$\sim 10^{-4}$~pc~cm$^{-3}$, which may be detectable with sensitive
pulse timing observations.
Additionally, the fits establish a joint constraint on the distance to
the NS and the ambient density, D$_{\rm kpc} = 0.48 n_{A}^{-1/4}$.  
We note that the densities obtained above are low, suggesting a possible
overestimate of the distance. For an ambient density of
0.05~cm$^{-3}$, which is comparable to the density of the local warm
ionized medium \citep[e.g.][]{P84}, the implied distance is 1~kpc
(and the height above the Galactic plane is reduced from 240~pc to 120~pc).

To check for unresolved small-scale structure in the limb-brightened
nebula, the autocorrelation function was calculated for sections of
the image with and without nebular emission. After accounting for
Poisson noise and the contribution from smooth extended structure, the
excess in the on-nebula autocorrelation function due to
barely-resolved or unresolved structure (50 mas or less) is $<5\%$ of
the larger-scale nebular emission.  We conclude that 50~mas represents
a lower limit on the angular scale of structure in the interstellar
medium probed by these HST observations of the Guitar nebula.


\section{Discussion}

\begin{figure}[ht]
\epsscale{1.0}
\plotone{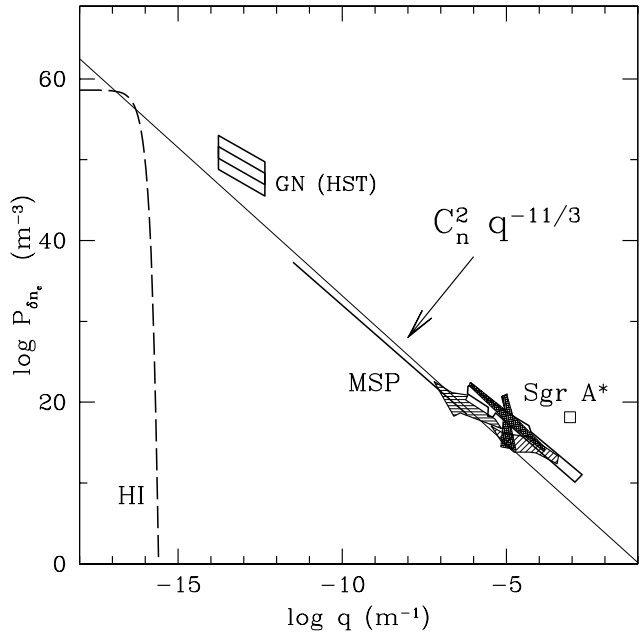}
\caption{Power spectrum for electron density variations in the ISM,
defined such that the integral over all wavenumbers is the mean-square
electron density.  The region labeled `GN' designates the constraints
derived from HST observations of the Guitar Nebula described here.
The wavenumber range corresponds to the reciprocals of the length
scales probed on angular scales between $\sim 0.05$ and $1.3$ arcsec,
assuming a distance of 1.9 kpc and a combined filling factor and
ionized fraction ranging from 0.02 to 0.5.  The dashed line labelled
`HI' is an estimate using the length scales of standard HI clouds
\citep{MO77}, assumed to be partially ionized.  The line labelled
`MSP' represents variations in dispersion measure for the millisecond
pulsar B1937+21 \citep{CWD+90,KTR94} The observational constraints at
the highest wavenumbers result from measurements of interstellar
scintillation and scattering.  The point for Sgr~A* represents the
power level for scattering of the Galactic center source, corrected
for the proximity of the scattering screen to the source \citep{LC98}.
The hatched regions incorporate constraints from scintillation and
scattering observed for several hundred sources \citep[NE2001;][and
references therein]{CL02}.  The light solid line designates a power
spectrum with power-law index equal to the Kolmogorov value ($-11/3$).
\label{fig-spec} 
} 
\end{figure}

The wavenumber spectrum for electron density fluctuations in the
general ISM has been delineated through a variety of measurements,
including radio scintillation, scattering, pulse time-of-arrival, 
and Faraday-rotation, as illustrated in Figure~\ref{fig-spec}.  
These measurements constrain the power levels in different wavenumber
intervals, and suggest an overall consistency with a Kolmogorov
turbulence process \citep{ARS95}, although this consistency is only
coarse and may be an illusion associated with the large amplitude
scale of the diagram.  Departures from the Kolmogorov spectrum are
also inferred from pulsar scintillation measurements.  The HST
measurements of the Guitar Nebula provide constraints on wavenumbers
not easily accessible by other probes and therefore provide a new tool
for investigating fine structure in the interstellar density.

Along with the properties of the ISM, the Guitar nebula also provides
a probe of the pulsar's relativistic wind.  The properties of pulsar
winds, including the magnetization parameter $\sigma$, the ratio of
the Poynting flux to the kinetic energy flux, have been inferred
primarily from the Crab nebula \citep{KC84,GA94,MM96}.  The existence
of bow shock nebulae such as the Guitar requires collisional
excitation of the neutral interstellar medium through interactions
with shocked electrons and protons. These charged particles can be
ejected from the NS itself, or originate from interstellar atoms
through photoionization or magnetic reconnection outside the pulsar
light cylinder radius ($R_{LC} = c P/2\pi$). The process by which the
NS Poynting flux is converted to particle kinetic energy is not well
understood: overviews of different processes are provided by
\citet{BL92}, \citet[ion loading]{conf-GVKA02} and \citet[striped
pulsar winds]{LK01}. At the stand-off radius ($R_0 \gg R_{LC}$),
pressure balance requires that the spindown energy of the NS be
carried by the particle flux ($\sigma\ll 1$), but the shape of the
nebula may encode information about the (rotation-averaged) shape of
the NS wind.

Along with the Guitar nebula, future observations of the evolution of
bow shock nebulae will be possible for the nearby radio-quiet neutron
star RX~J1856.5$-$3754 \citep{VK01}, only $\sim$120 pc away
\citep{KVA02,WL02}.  Even at its relatively low speed ($<
200$~km~s$^{-1}$), the neutron star travels $\sim 1$\arcsec\ in a
year, and evolution of the nebula should be evident on roughly this
time scale.  PSR~J2124$-$3358, a nearby millisecond pulsar with a
complex bow shock nebula \citep{GJS02}, is also promising in this
regard, while the discovery of a bow shock nebula powered by the
Poynting flux from a magnetar would allow investigation of the
relativistic wind in the presence of an ultra-strong magnetic field.

\acknowledgements 

This work is based in part on observations made with the NASA/ESA
Hubble Space Telescope, obtained at the Space Telescope Science
Institute, which is operated by the Association of Universities for
Research in Astronomy, Inc., under NASA contract NAS 5-26555.  These
observations are associated with proposals 5387 and 9129.  This work
was also supported by NSF grants AST 9819931 and AST 0206036, and made
use of NASA's Astrophysics Data System Abstract Service and the {\tt
arXiv.org} astro-ph preprint service.

\end{document}